\documentclass[a4paper,english]{lipics-v2019}
\usepackage{breqn}
\usepackage{ifthen}
\usepackage{placeins}
\usepackage{complexity}
\usepackage{nicefrac}
\usepackage{todonotes}
\usepackage{siunitx}
\usepackage{complexity}
\usepackage{amssymb}
\usepackage{color}
\usepackage[noabbrev,capitalise]{cleveref}
\usepackage{mathtools}

\hideLIPIcs
\nolinenumbers

\newcommand{\old}[1]{{}}

\title{Central Triangulation under Parallel Flip Operations:\\ The CG:SHOP Challenge 2026}
\titlerunning{Minimum Parallel Flip Distance: CG Challenge 2026}

\author{Oswin Aichholzer}{Institute of Algorithms and Theory, TU Graz, Austria}{oswin.aichholzer@tugraz.at}{https://orcid.org/0000-0002-2364-0583}{}
\author{Joseph Dorfer}{Institute of Algorithms and Theory, TU Graz, Austria}{joseph.dorfer@tugraz.at}{https://orcid.org/0009-0004-9276-7870}{Austrian Science Fund (FWF) 10.55776/DOC183.}
\author{Sándor P.~Fekete}{Department of Computer Science, TU Braunschweig, Germany}{s.fekete@tu-bs.de}{https://orcid.org/0000-0002-9062-4241}{}
\author{Phillip Keldenich}{Department of Computer Science, TU Braunschweig, Germany}{p.keldenich@tu-bs.de}{https://orcid.org/0000-0002-6677-5090}{}
\author{Peter~Kramer}{Department of Computer Science, TU Braunschweig, Germany}{kramer@ibr.cs.tu-bs.de}{https://orcid.org/0000-0001-9635-5890}{}
\author{Dominik Krupke}{Department of Computer Science, TU Braunschweig, Germany}{d.krupke@tu-bs.de}{https://orcid.org/0000-0003-1573-3496}{}
\author{Stefan Schirra}{Department for Simulation and Graphics, OvGU Magdeburg, Germany}{stschirr@isg.cs.uni-magdeburg.de}{https://orcid.org/0009-0006-5928-1494}{}
\authorrunning{O.~Aichholzer, J.~Dorfer, S.~P.~Fekete, P.~Keldenich, P.~Kramer, D.~Krupke, S.~Schirra}
\Copyright{O.~Aichholzer, J.~Dorfer, S.~P.~Fekete, P.~Keldenich, P.~Kramer, D.~Krupke, S.~Schirra}



\supplement{}

\funding{Work at TU Braunschweig was partially supported by the German Research Foundation (DFG),
    project ``Computational Geometry: Solving Hard Optimization Problems'' (CG:SHOP), grant 444569951.}

\acknowledgements{We thank the members of the CG:SHOP Challenge Advisory Board for their valuable input:
William J.~Cook, Andreas Fabri, Dan Halperin, Michael Kerber, Philipp Kindermann, Joe Mitchell, Kevin Verbeek.}

\EventAcronym{CG:SHOP 2026}

\begin{document}
    \maketitle
    \begin{abstract}
        We give an overview of the 2026 Computational Geometry Challenge targeting the problem of finding a \textcolor{black}{\textsc{Central Triangulation under Parallel Flip Operations} in triangulations of point sets.
        A \emph{flip} is the parallel exchange of a set of edges in a triangulation with opposing diagonals of the convex quadrilaterals containing them.
        The challenge objective was, given a set of triangulations of a fixed point set, to determine a \emph{central} triangulation with respect to parallel flip distances.
        More precisely, this asks for a triangulation that minimizes the sum of flip distances to all elements of the input}.

    \end{abstract}

    \section{Introduction}
    The ``CG:SHOP Challenge'' (Computational Geometry: Solving Hard
    Optimization Problems) originated as a workshop at the 2019
    Computational Geometry Week (CG Week) in Portland, Oregon in June,
    2019. The goal was to conduct a computational challenge competition
    that focused attention on a specific hard geometric optimization
    problem, encouraging researchers to devise and implement solution
    methods that could be compared scientifically based on how well they
    performed on a database of carefully selected and varied instances.
    While much of computational
    geometry research is theoretical, often seeking provable approximation
    algorithms for \NP-hard optimization problems,
    the goal of the Challenge was to set the metric of success based on
    computational results on a specific set of benchmark geometric
    instances. The 2019 Challenge~\cite{CGChallenge2019_JEA} focused on the problem of computing
    simple polygons of minimum and maximum area for given sets of vertices in the
    plane. It generated a strong response from many research
    groups~\cite{area-crombez,area-tau,area-salzburg,area-exact,area-campinas,area-omega} from both the computational geometry and the combinatorial
    optimization communities, and resulted in a lively exchange of
    solution ideas.

    Subsequently, the CG:SHOP Challenge became an event within the CG Week
    program, with top performing solutions reported in the Symposium on
    Computational Geometry (SoCG) proceedings.
    The schedule for the Challenge was
    advanced earlier, to give an opportunity for more participation, particularly
    among students, e.g., as part of course projects.
    For CG Weeks 2020, 2021, 2022, 2023, 2024, 2025
    the Challenge problems were \textsc{Minimum Convex Partition}~\cite{CGChallenge2020,SoCG2020_1,SoCG2020_2,SoCG2020_3},
    \textsc{Coordinated Motion Planning}~\cite{CGChallenge2021,Challenge2021_1,Challenge2021_2,Challenge2021_3,SoCG2021_1,SoCG2021_2,SoCG2021_3},
    \textsc{Minimum Partition into Plane Subgraphs}~\cite{CGChallenge2022_JEA,Challenge2022_1,Challenge2022_2,SoCG2022_1,SoCG2022_2,SoCG2022_3,SoCG2022_4},
    \textsc{Minimum Convex Covering}~\cite{fekete2023minimum,Challenge2023_1,Challenge2023_2,SoCG2023_1,SoCG2023_2},
    \textsc{Maximum Polygon Packing}~\cite{fekete2024maximum,Challenge2024_1,Challenge2024_2,Challenge2024_3,Challenge2024_4},
    and
    \textsc{Minimum Non-Obtuse Triangulations}~\cite{fekete2025minimum,Challenge2025_1,Challenge2025_2},
    respectively. Hosting the competition is a challenge in and of itself; see~\cite{Challenge2023_0} for a detailed
    exposition.

    The eighth edition of the Challenge in 2026 continued
    this format, leading to contributions in the SoCG proceedings.

    \section{The Challenge: Central Triangulation under Parallel Flip Operations}

    A suitable contest problem has a number of desirable properties.

    \begin{enumerate}
        \item The problem is of geometric nature.
        \item The problem is of general scientific interest and has received previous attention.
        \item Optimization problems tend to be more suitable than feasibility problems; in principle,
        feasibility problems are also possible, but they need to be suitable for sufficiently
        fine-grained scoring to produce an interesting contest.
        \item Computing optimal solutions is difficult for instances of reasonable size.
        \item This difficulty is of a fundamental algorithmic nature, and not only due to
        issues of encoding or access to sophisticated software or hardware.
        \item Verifying feasibility of provided solutions is relatively easy.
    \end{enumerate}

    In this eighth year, a call for suitable problems was communicated in April
    2025. In the end, the selected problem was chosen, which had recently gained
    wider interest in the community and received considerable attention, mostly 
    on the theoretical side. Due to the outstanding expertise on the topic
    (among other things, illustrated by a well-received survey talk at EuroCG 2025)
    the Challenge team was joined by experts on the topic: Oswin Aichholzer, Joseph Dorfer,
    and Peter Kramer.

    \subsection{The Problem}
    \textcolor{black}{The specific problem that formed the basis of the 2026 CG Challenge was based on flip operations in triangulations of point sets in the plane.
    In particular, a \emph{flip} in a triangulation~$T$ of a point set $P$ exchanges an edge $e$ that is the diagonal of a convex quadrilateral of $T$ (i.e., $e$ is a flippable edge) by the other diagonal of this convex quadrilateral in order to obtain a new triangulation $T'$.
    By extension, a \emph{parallel flip} in a triangulation $T$ simultaneously flips a set~$D$ of diagonals of convex quadrilaterals in $T$, such that no two edges $e,e' \in D$  share a triangle.
    The \emph{parallel flip distance} $\delta(T,T')$ denotes the minimum number of parallel flips necessary to obtain one triangulation from the other, that is, the length of a shortest parallel flip sequence.
    See~\cref{fig:flips} for illustrations of both types of flip operation.
        \begin{figure}[hbt]%
            \captionsetup[subfigure]{justification=centering}%
            \begin{subfigure}[t]{0.5\textwidth-5pt}%
                \centering%
                \includegraphics[page=2]{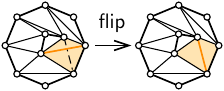}%
                \caption{}%
                \label{fig:flip}
            \end{subfigure}%
            \hfill%
            \begin{subfigure}[t]{0.5\textwidth-5pt}%
                \centering%
                \includegraphics[page=3]{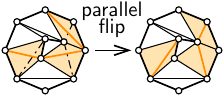}%
                \caption{}%
                \label{fig:parallel-flip}
            \end{subfigure}
            \caption{Illustrative examples of (a) sequential and (b) parallel flip operations.}%
            \label{fig:flips}%
        \end{figure}
    }

\begin{figure}[tbh]%
        \centering%
        \includegraphics[page=1]{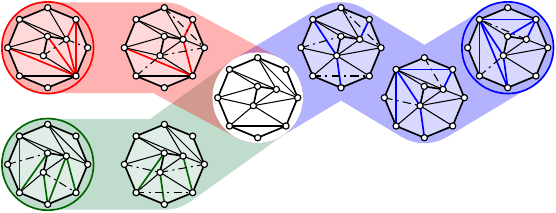}%
        \caption{Three triangulations $T_1,T_2,T_3$ and a central triangulation $C$ with objective value $7$.}%
        \label{fig:central-triangulation-example}%
    \end{figure}

    \textcolor{black}{
        \medskip
        \begin{description}
            \item[{Problem:}] \textsc{Central Triangulation under Parallel Flip Operations}
            \item[{Given:}] A set of $m$ triangulations $T_1,\ldots,T_m$ on the same set $P$ of $n$ points in the plane.
            \item[{Goal:}] Determine a triangulation $C$ on $P$, such that the sum of parallel flip distances from $C$ to the input triangulations $\sum_{i=1}^m\delta(C,T_i)$ is minimized. See~\cref{fig:central-triangulation-example} for an example.
        \end{description}
        Flipping problems, particularly in geometric triangulations, have a long history within computational geometry. For a comprehensive overview, see~\cref{sec:related}.
    }

    \subsection{Scoring}

    For each instance, points were awarded based on the rank of the submitted solution relative to other teams.
    Equal performance yielded equal points, and the overall score was the sum of points across all instances.
    The best solution for an instance earned 40 points, the second-best 32, the third-best 25, and so on, as shown in~\cref{tab:scoring}.
    If multiple teams tied for a rank, all received the same points, and the next distinct solution was assigned the points corresponding to the subsequent rank.
    For example, if two teams shared the best objective value, both received 40~points, and the next distinct solution received 25~points (third-place points).
    In case of ties in overall score, the tiebreaker was the time at which a specific score was first obtained; participants were encouraged to submit early and often.

    This scoring scheme rewarded both optimal solutions for smaller instances and scalable heuristics for larger ones, for which different techniques may be required.
    Even small improvements could result in noticeably higher scores, allowing exact methods to earn substantial points on smaller instances, while compensating for their limited scalability on larger ones.

    \begin{table}[ht]
        \centering
        \begin{tabular}{lcccccccccccc}
            \hline
            \textbf{Rank} & 1 & 2 & 3 & 4 & 5 & 6 & 7 & 8 & 9 & 10 & 11 & 12 \\
            \textbf{Points} & 40 & 32 & 25 & 19 & 14 & 10 & 7 & 5 & 4 & 3 & 2 & 1 \\
            \hline
        \end{tabular}
        \caption{Points awarded per instance based on solution rank.}
        \label{tab:scoring}
    \end{table}

    \subsection{Related Work} \label{sec:related}

    Flips in triangulations have been widely studied when the
    flips are performed \emph{sequentially}.
    Lawson~\cite{lawson1972transforming} showed
    that any triangulation on a point set in general position in the plane can be transformed into any other in $\mathcal{O}(n^2)$ flips.
    Hurtado, Noy, and Urrutia~\cite{hurtado1996flipping} provided a matching lower bound by
    constructing a pair of triangulations for which any flip sequence requires $\Omega(n^2)$ steps to transform one triangulation into the other.
    Hanke, Ottmann, and Schuierer~\cite{hanke1996edge} showed that the number of crossings between the edges of two triangulations is an upper bound on the flip distance of the two triangulations.

    Another central question is the complexity of the
    \emph{flip distance problem}, which asks to determine the minimum number of flips required to transform one triangulation into another.
    \textcolor{black}{An important tool for the investigation of this question are \emph{happy edges}.
    An edge is \emph{happy} in an instance of the flip distance problem exactly if it is contained in both the start and target triangulation.
    The \emph{happy edge property}, in the context of any edge-flipping problem, holds if there always exists a shortest flip sequence which does not flip such edges at all.
    Determining this for any given model can give deep insight into the structure of shortest flip sequences.}

    Eppstein~\cite{eppstein2007happy} provided a polynomial-time algorithm to determine the flip distance of triangulations on point sets, not necessarily in general position, without empty 5-holes.
    Lubiw and Pathak~\cite{lubiw2015flip} and Pilz~\cite{pilz2014flip} independently proved that the flip distance problem is \NP-hard for triangulations of a point set in general position\textcolor{black}{, building on counterexamples to the happy edge property}.
    The former paper also proves \NP-hardness for the flip distance problem for triangulations of polygons with holes.
    The latter paper proves \APX-hardness of the flip distance of triangulations of point sets in general position, ruling out the existence of a \PTAS.
    Aichholzer, Mulzer, and Pilz~\cite{aichholzer2015flip} proved \NP-hardness for the flip distance problem for triangulations of simple polygons.
    Most recently, Dorfer~\cite{dorfer2026flip} showed that the flip distance problem is already \NP-hard for triangulations of convex polygons.
    This problem was of particular interest because it is equivalent to determining the rotation distance between binary trees.
    In parallel to these hardness results, a series of \FPT~algorithms for the computation of the flip distance $k$, when parametrized by $k$, \textcolor{black}{emerged~\cite{feng2021improved,kanj2017computing}, with the current best running time being~$\mathcal{O}(n+k32^k)$~\cite{feng2021improved}.}
	Particularly notable on the \emph{practical} side is the work by Mayer and Mutzel~\cite{mayer2024engineering}, who used an A* star algorithm for computing optimal flip sequences between pairs of triangulations.

    \textcolor{black}{
    Significantly less is known about parallel flip operations.
    Hurtado, Noy, and Urrutia~\cite{hurtado1998parallel} first proved that $\Omega(n)$ parallel flips between two triangulations of $n$ points are sometimes necessary, and gave an upper bound of $\mathcal{O}(n\log n)$ on the parallel flip distance, which was later improved by the authors of~\cite{galtier2003simultaneous} to $\mathcal{O}(n)$.
    The same publication shows that at least $\frac{n-4}{5}$ edges can be flipped in parallel in any triangulation, which has since been proven to be tight by Souvaine, Tóth, and Winslow~\cite{souvaine2012simultaneous}.
    To the best of our knowledge, it is not known whether the happy edge property holds for parallel flips in triangulations of general point sets.
    }

    \subsection{Instances}

    The generation of appropriate instances is critical for any challenge.
    Instances that are too easily solved undermine the complexity of the problem, rendering it trivial.
    Conversely, instances that require significant computational resources for preprocessing or for finding viable solutions may offer an unfair advantage to teams with superior computing facilities.
    This issue is amplified when the instance set is so large that it exceeds the management capacity when using a limited number of computers.

    An initial set of \num{201} small to medium instances from two classes (\emph{random} and \emph{woc}) was released on October~15, 2025.
    Shortly after, a participant from the eventual winning team (Shadoks) pointed out that the initial instances might be too small to sufficiently differentiate between strong teams, noting that exact parallel flip distances for instances with hundreds of points could be computed in seconds using a SAT solver, and that heuristics performed well even for the center problem.
    In response, \num{49} considerably larger instances from a third class (\emph{rirs}) were added on October~27, with up to \num{12500} points and 200 triangulations, to ensure that scalable heuristics would be necessary for competitive performance.

    The final benchmark set comprised \num{250} instances from three distinct classes:
    \begin{description}
        \item[random] \num{100} instances with 15--320 points and 2--20 triangulations.
        \item[woc] \num{101} instances with 45--250 points and 3--7 triangulations.
        \item[rirs] \num{49} large-scale instances with 500--\num{12500} points and 20--200 triangulations.
    \end{description}

    \subsubsection{Random instances}
    The \emph{random} instances were generated by first sampling point sets from TSPLIB~\cite{reinelt1991tsplib}, a well-known collection of Traveling Salesman Problem instances with integer coordinates.
    For each instance, one of 29~curated TSPLIB instances was chosen at random, and a subset of its points was sampled without replacement.
    This yields realistic, non-uniform point distributions rather than uniformly random placements.

    An initial \emph{center} triangulation was then constructed by considering all possible edges in random order and greedily inserting each edge that did not intersect any previously selected edge.
    From this center, each of the $m$~target triangulations was obtained independently via a random parallel flip walk:
    in each of \texttt{num\_steps} rounds, a maximal independent set of flippable edges was identified, and each flip was applied with a fixed probability~\texttt{prob}.
    \begin{figure}
	\centering
	\includegraphics[width=.99\textwidth]{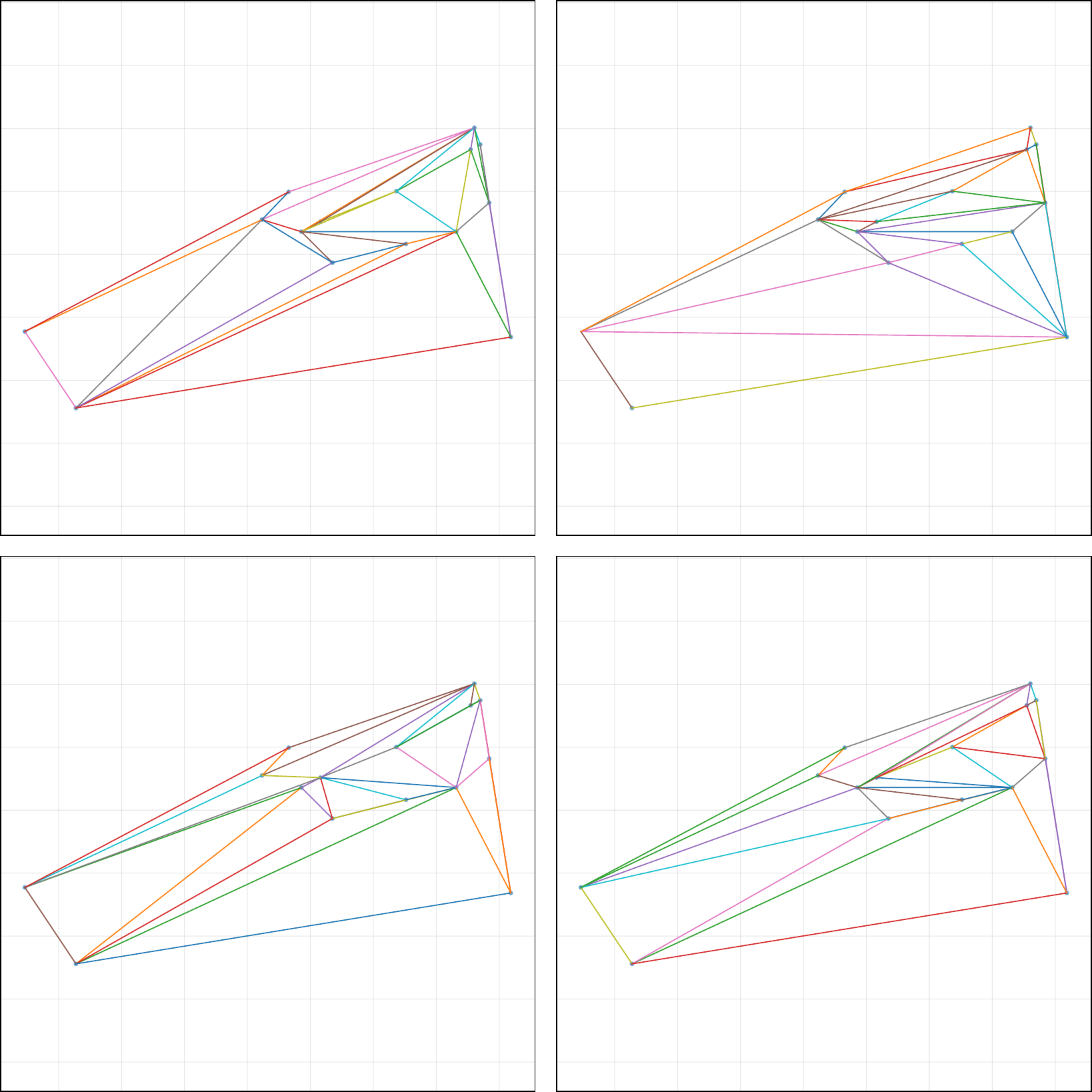}
	\caption{An example \emph{random} instance with \num{15} points and \num{4} triangulations for which a center is to be found.}
    \end{figure}
    The generation parameters were varied across instances, with \texttt{num\_steps} $\in \{10, 20, 50, 100, 1000\}$ and \texttt{prob} $\in \{0.1, 0.3, 0.5, 0.7, 0.9\}$, yielding triangulations with varying degrees of similarity to the center.
    Notably, the center triangulation provides a known feasible solution whose objective value serves as an upper bound on the optimum.

    \subsubsection{Well-known Objective Collection: The woc instances}
\begin{figure}
	\centering
	\includegraphics[width=.99\textwidth]{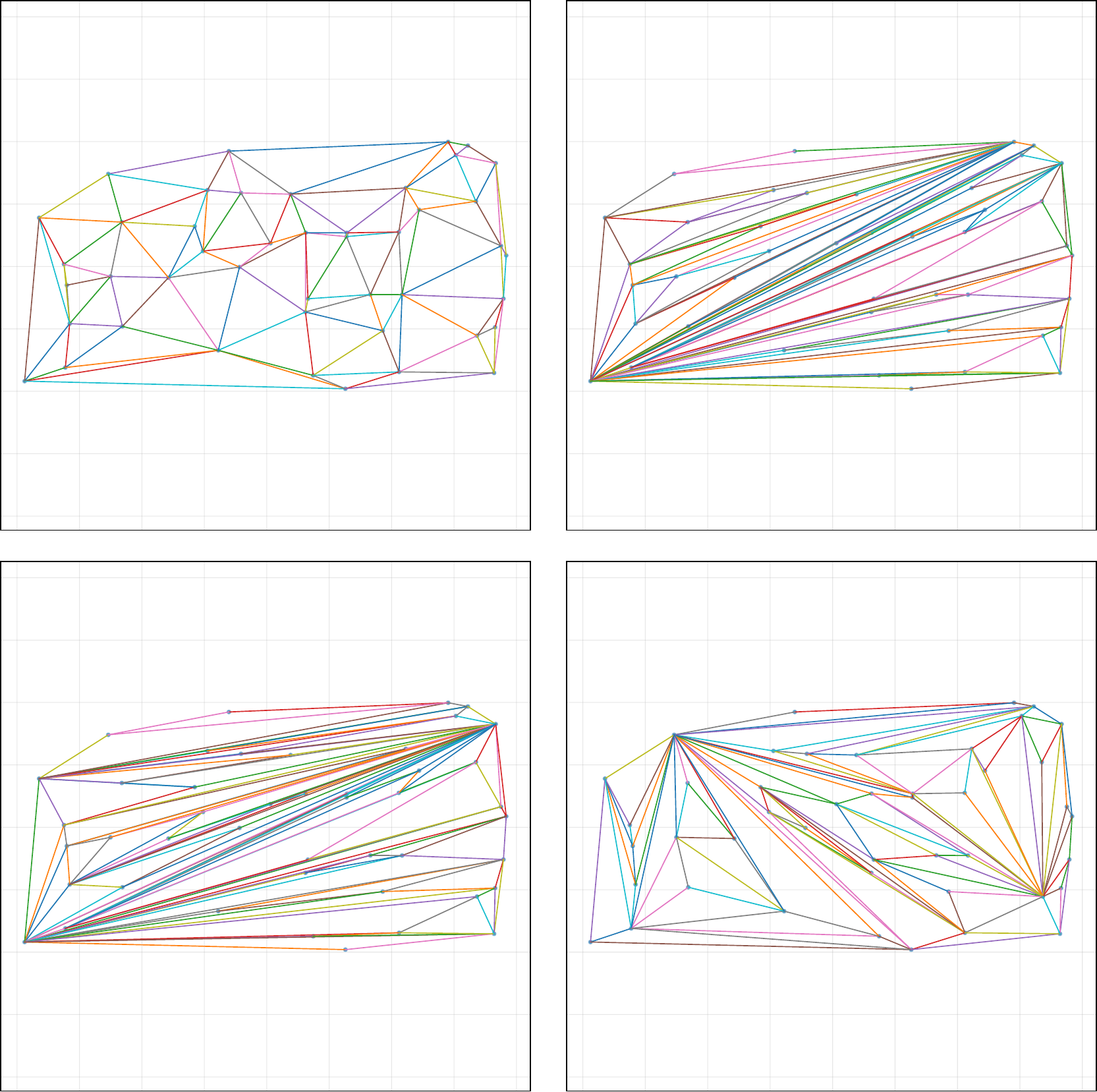}
	\caption{An example \emph{woc} instance with \num{45} points and \num{4} triangulations for which a center is to be found.}
	\label{fig:example_woc}
    \end{figure}

    The \emph{woc} instances were designed so that the input triangulations are structurally diverse by construction.
    For each instance, a point set was taken either from TSPLIB~\cite{reinelt1991tsplib} or generated randomly, and then multiple triangulations were computed by optimizing different objective functions on the same point set.
    This included the Delaunay triangulation, the Minimum Weight Triangulation (MWT), Maximum Weight Triangulation, Minimum Dilation Triangulation, as well as triangulations optimizing max min edge length, min max edge length, 
    min max triangle area, and max min triangle area criteria; see \cref{fig:example_woc} for an example.
    Because some of the solvers used to compute these triangulations ran out of time for larger point sets or resulted in (almost) identical triangulations, not all of these objectives are present in all woc instances.
    Additionally, constrained Delaunay triangulations with random constraints were used.
    Because triangulations optimizing different objectives tend to differ substantially in structure, this yields instances where the input triangulations have relatively low pairwise edge overlap, 
    hopefully making the search for a good central triangulation more challenging despite the moderate instance sizes (45--250 points).

    \subsubsection{Random Insertion, Random Shuffle: The rirs instances}
    The \emph{rirs} instances were the largest in the benchmark set and were added after the initial release to ensure that scalable approaches were necessary.
    Point sets were generated uniformly at random with integer coordinates.
    Unlike the \emph{random} class, each triangulation was generated \emph{independently} by considering all possible edges in a random order and greedily inserting each edge that did not intersect any previously selected edge.
    Because each triangulation uses a fresh random permutation of edges, the resulting triangulations are only weakly correlated, with pairwise edge overlap around 20\%---substantially lower than in the flip-walk-based \emph{random} instances.
    With up to \num{12500} points and \num{200} triangulations, these instances were specifically designed to test the scalability of participants' approaches.

    \subsection{Categories}

    The contest was run in an \emph{Open Class}, in which participants could use any
    computing device, any amount of computing time (within the duration of the
    contest) and any team composition.
    To be eligible for the \emph{Junior Class}, a team had to consist exclusively of participants who were eligible according to the rules of CG:YRF (the \emph{Young Researchers Forum} of CG Week), defined as not having defended a formal doctorate before 2024. Additionally, a team could include one senior academic advisor, who was allowed to provide guidance and assist with writing the final paper (if the team was selected for inclusion in the proceedings), but all programming tasks had to be performed by the junior team members.

    \subsection{Server and Timeline}

    The competition was facilitated through a dedicated server at TU Braunschweig,
    accessible at \url{https://cgshop.ibr.cs.tu-bs.de/competition/cg-shop-2026/}.
    An initial batch of test instances was made available on September~15, 2025,
    followed by the release of \num{201} small to medium benchmark instances on October~15, 2025.
    In response to community feedback, an additional 49 large instances were released on October~27, 2025.
    The competition concluded on January~29, 2026 (AoE).

    Participants were provided with a verification tool as an open-source Python package, available at \url{https://github.com/CG-SHOP/pyutils26}.
    This tool provided detailed information on errors and allowed participants to easily investigate and correct their solutions if the server rejected them.

    \section{Outcomes}

    A total of 18 teams submitted at least one valid solution.
    \Cref{tab:team_ranks} shows the final rankings.
    The first-place team, Shadoks, achieved a score of \num{9992} out of a theoretical maximum of \num{10000} (so they scored the maximum of 40~points on all but one of the 250~instances, where they came in second), demonstrating near-perfect dominance across all instances.
    The second-place team, ETH Flippers, was also recognized as the best junior team with a score of \num{8474}.

    \begin{table}[ht]
        \centering
        \begin{tabular}{rlrcrr}
            \hline
            \textbf{Rank} & \textbf{Team Name} & \textbf{Score} & \textbf{Junior} & \textbf{\# best} & \textbf{\# sole best} \\
            \hline
            1  & Shadoks                       & \num{9992} &              & 249 & 56 \\
            2  & ETH Flippers                  & \num{8474} & \checkmark   & 186 &  1 \\
            3  & CG\#Hunters                   & \num{7777} &              & 103 &  0 \\
            4  & jflip2                        & \num{7764} &              & 114 &  0 \\
            5  & TU Dortmund                   & \num{6265} &              &  81 &  0 \\
            6  & UCPH                          & \num{3100} & \checkmark   &  41 &  0 \\
            7  & Anonymous                     & \num{2840} &              &  29 &  0 \\
            8  & misojiro                      & \num{2612} & \checkmark   &  28 &  0 \\
            9  & Too square to triangle        & \num{2451} & \checkmark   &  32 &  0 \\
            10 & Anonymous                     &   \num{953} & \checkmark   &  14 &  0 \\
            11 & SWP TI FU Berlin Team T       &   \num{923} & \checkmark   &  12 &  0 \\
            12 & UMass Lowell ComGeo Group     &   \num{887} &              &  10 &  0 \\
            13 & KITIS                         &   \num{532} &              &  12 &  0 \\
            14 & Team Consisting of Nick Belov &   \num{380} & \checkmark   &   0 &  0 \\
            15 & team Marseille                &   \num{124} & \checkmark   &   2 &  0 \\
            16 & t.mamashloo                   &   \num{120} &              &   3 &  0 \\
            17 & Anonymous                     &    \num{80} & \checkmark   &   2 &  0 \\
            18 & The Triangulators             &     \num{8} & \checkmark   &   0 &  0 \\
            \hline
        \end{tabular}
        \caption{Team rankings with scores and junior team status. \textbf{\# best} denotes the number of instances where the team achieved the best known objective value. \textbf{\# sole best} counts instances where the team was the only one to achieve the best value.}
        \label{tab:team_ranks}
    \end{table}

As a result, the top four teams were invited to contribute to the 2026 SoCG proceedings:

\begin{enumerate}
\item 
Team Shadoks: Guilherme da Fonseca, Fabien Feschet, and Yan Gerard~\cite{Challenge2026_1}.
\item
Team ETH Flippers: Lorenzo Battini and Marko Milenkovi\'c~\cite{Challenge2026_2}
\item
Team CG\#Hunters: Jaegun Lee, Seokyun Kang, Hyeonseok Lee, Hyeyun Yang, and Taehoon Ahn~\cite{Challenge2026_3}
\item
Team jflip2: Jacobus Conradi, Benedikt Kolbe, Philip Mayer, Jonas Sauer, and Jack Spalding-Jamieson~\cite{Challenge2026_4}
\end{enumerate}

Even within this top group, a variety of different approaches was used, with the
top teams using both exact and heuristic methods.
The approach by Shadoks (the dominant frontrunner) combines exact computation based
on SAT with several greedy heuristics and also makes use of SAT and MaxSAT for
solution improvement.  ETH Flippers used an exact solver for small and medium-sized instances
and a suite of heuristics for larger instances. CG\#Hunters
focused on heuristic approaches with local improvement, utilizing
initial heuristic solutions with greedy and large neighborhood search.
Team jflip2 combined a heuristic for initial solutions with local improvement
in a simulated-annealing framework.

\begin{figure}[tb]
        \centering
        \includegraphics[width=\textwidth]{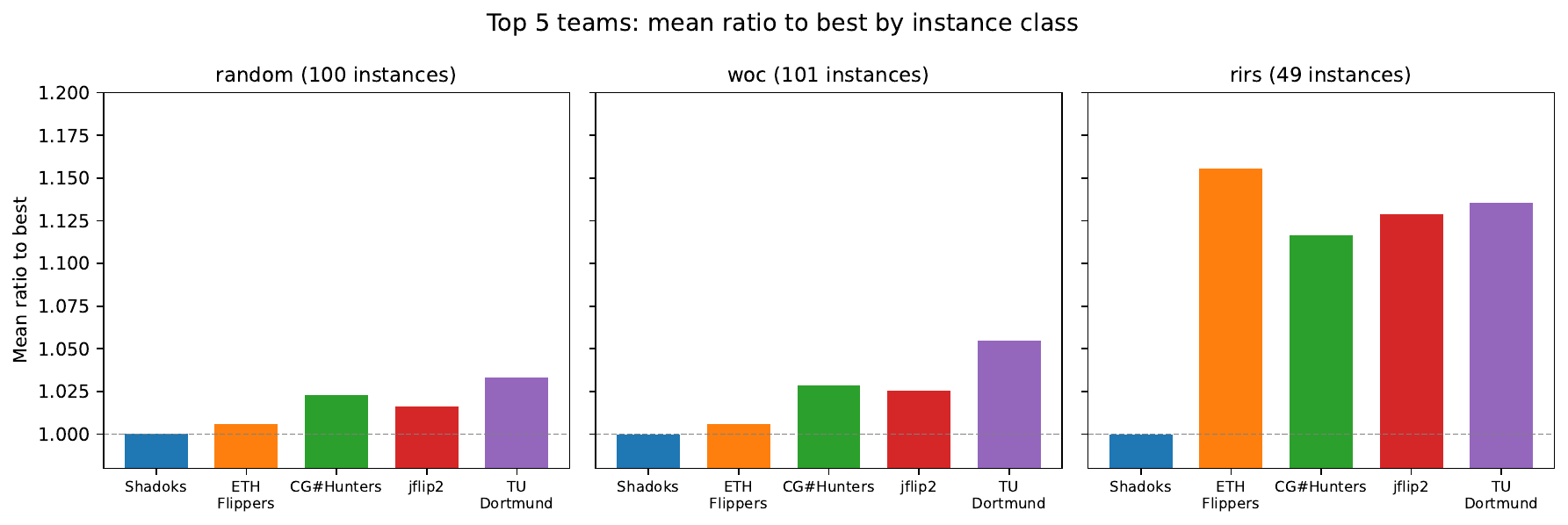}
        \caption{Mean ratio to the best solution for the top~5 teams, by instance class. A value of 1.0 means the team matched the best known solution on every instance in that class.}
        \label{fig:top5_ratio_by_class}
    \end{figure}

    \Cref{fig:top5_ratio_by_class} shows the mean ratio to the best solution for each of the top five teams, broken down by instance class.
    On the \emph{random} and \emph{woc} instances, all five teams performed within a few percent of each other, with Shadoks achieving the best solution on nearly every instance.
    The \emph{rirs} instances revealed the largest differences: While Shadoks remained at 1.0, the remaining teams were 12--16\% worse on average.
    Notably, the ranking among teams 2--5 shifts between instance classes: ETH~Flippers was strongest on \emph{random} and \emph{woc}, but weakest of the four on \emph{rirs}, where CG\#Hunters consistently held second place.

\begin{figure}[tb]
        \centering
        \begin{subfigure}[t]{0.49\textwidth}
            \centering
            \includegraphics[width=\textwidth]{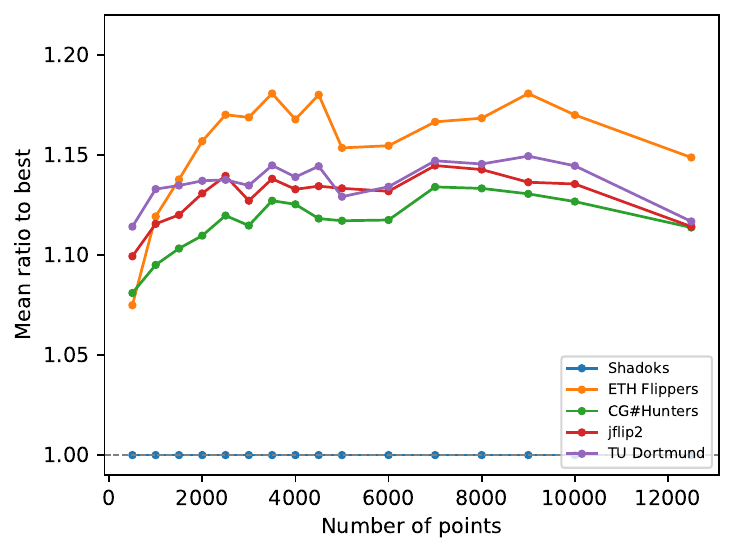}
            \caption{Ratio to best vs.\ number of points.}
            \label{fig:top5_rirs_scaling}
        \end{subfigure}
        \hfill
        \begin{subfigure}[t]{0.49\textwidth}
            \centering
            \includegraphics[width=\textwidth]{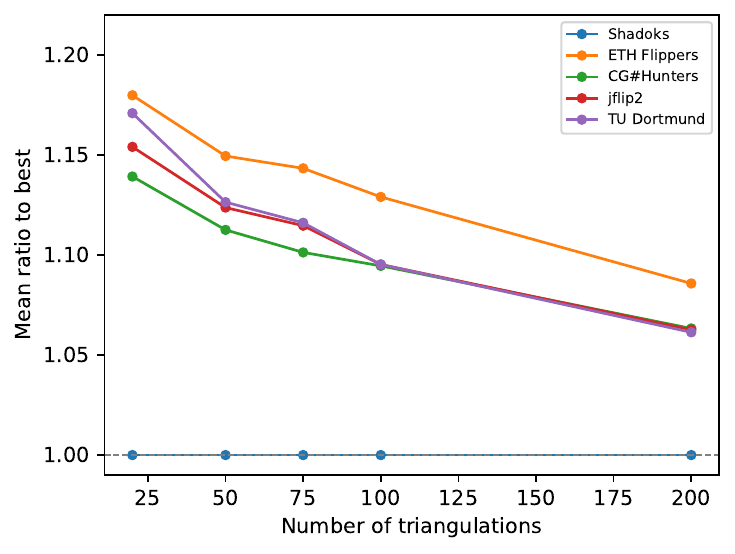}
            \caption{Ratio to best vs.\ number of triangulations.}
            \label{fig:top5_rirs_by_triangulations}
        \end{subfigure}
        \caption{Mean ratio to best on \emph{rirs} instances for the top~5 teams. (a)~The gap stabilizes beyond roughly \num{2000} points. (b)~More triangulations lead to smaller performance gaps.}
        \label{fig:top5_rirs}
    \end{figure}

    \Cref{fig:top5_rirs} provides a closer look at the \emph{rirs} instances.
    \Cref{fig:top5_rirs_scaling} shows how performance varies with instance size: the gap between Shadoks and the other teams stabilizes after roughly \num{2000} points, suggesting that all top teams use similarly scaling algorithms, but that Shadoks has a constant-factor advantage.
    \Cref{fig:top5_rirs_by_triangulations} shows the effect of the number of triangulations.
    With more triangulations, the gap between teams narrows considerably---from 14--18\% at $m=20$ to 6--9\% at $m=200$---suggesting that a larger number of input triangulations constrains the solution space more tightly, leaving less room for algorithmic differences.

    \section{Conclusions}
    The 2026 CG:SHOP Challenge motivated a considerable number of teams to engage
    in extensive optimization studies. The outcomes promise further insight into
    the underlying, important optimization problem. This demonstrates the
    importance of considering geometric optimization problems from a practical
    perspective.

    \bibliographystyle{plainurl}
    \bibliography{bibliography,references,references2026}
\end{document}